# Role of the Subunits Interactions in the Conformational Transitions in Adult Human Hemoglobin: an Explicit Solvent Molecular Dynamics Study.

Olaniyi K. Yusuff,[1,2,3] Jonathan O. Babalola,[1] Giovanni Bussi,*[3] and Simone Raugei[4]
[1.] Department of Chemistry, University of Ibadan, Ibadan, Nigeria.
[2.] Department of Chemistry, Lagos State University, Ojo, Lagos, Nigeria.
[3.] Statistical Biophysics Sector, International School for Advance Studies (SISSA) Trieste, Italy.
[4.] Chemical and Material Science Division, Pacific Northwest National laboratory, Richland, Washington 99354, United States.
*To whom correspondence may be addressed, e-mail: bussi@sissa.it.

**ABSTRACT**

Hemoglobin exhibits allosteric structural changes upon ligand binding due to the dynamic interactions between the ligand binding sites, the amino acids residues and some other solutes present under physiological conditions. In the present study, the dynamical and quaternary structural changes occurring in two unligated (deoxy-) T structures, and two fully ligated (oxy-) R, R2 structures of adult human hemoglobin were investigated with molecular dynamics. It is shown that, in the sub-microsecond time scale, there is no marked difference in the global dynamics of the amino acids residues in both the oxy- and the deoxy- forms of the individual structures. In addition, the R, R2 are relatively stable and do not present quaternary conformational changes within the time scale of our simulations while the T structure is dynamically more flexible and exhibited the T→R quaternary conformational transition, which is propagated by the relative rotation of the residues at the $\alpha_1\beta_2$ and $\alpha_2\beta_1$ interface.

**Keywords:** hemoglobin, allosteric regulation, molecular dynamics, quaternary conformational transition

## INTRODUCTION

Hemoglobin (Hb) is the iron- containing, oxygen-transporting metalloprotein in the red blood cells of animals. It transports oxygen from the lungs to the rest of the body where it releases the oxygen for cell use. The adult human hemoglobin (HbA) is a tetrameric protein molecule comprising of two α and two β subunits. The α and the β subunits are structurally identical consisting of 141 and 146 amino acids residues respectively. Each subunit contains a heme group at the centre to which molecular oxygen or other ligands bind. The heme group is linked to the protein part of the subunit by a covalent bond from the iron atom to a histidine F8 residue known as the proximal histidine. The oxygenation of HbA is regulated by allosteric mechanism which involves the interactions between the oxygen-binding sites (homotropic interactions) and interactions between individual amino acid residues in the protein molecule and various solutes which include hydrogen ions, chloride ions, carbon dioxide, inorganic phosphate ions, and organic phosphate ions, such as, 2,3-bisphosphoglycerate (2,3-BPG) and inositol hexaphosphate (IHP), (heterotropic interactions) (1). X-ray crystallography and other experimental studies have shown that the oxygen binding in Hb induces a quaternary structural transition from a low oxygen affinity deoxy- or unligated structure to a high oxygen affinity oxy- or fully ligated-structure. The unligated structure represents the 'tensed' or T-state while the fully ligated structure represents the 'relaxed' or R-state (2-5). Analytical studies of the x-ray structures of Hb in the unligated T- and the fully ligated R- states reveal that the tetramer maintains the same $\alpha_1\beta_1$ and $\alpha_2\beta_2$ interfaces but differs greatly at the $\alpha_1\beta_2$ and $\alpha_2\beta_1$ interfaces between the two dimers. Also a network of intra- and inter-subunits linkages (salt bridges) present at the amino- and carboxy- termini of $\alpha$- and $\beta$-chains in the unligated T-structure break on ligand binding. The regions where significant structural changes occur between the unligated and the ligated Hb include the hemes, the F helices, the FG corner regions of both subunits, and the E helix of the $\beta$ subunit (3,4,6). For a long time, the quaternary conformation of HbA was thought to exist in two forms: the unligated T-state and the fully ligated R-state. However, a third quaternary conformation for HbA representing a second fully ligated x-ray structure R2 has been found by Silva *et. al* (7) under low-salt and pH conditions. The structural analysis of the T, R and R2 conformations shows that the R-form lies on the transition path from the T-form to R2-form (8). Also, Safo and Abraham (9) reported two other x-ray structures for the fully ligated HbA, RR2 and R3, which were obtained under high-salt crystallization conditions of Perutz (10). They found that by rigid-body screw rotation, the RR2 structure was an intermediate conformation between R and R2 structures whereas the R3 structure exhibits a large quaternary structural difference from T structure as large as that of the R2, but in a different direction. NMR studies (11-13) have shown that CO ligated form of HbA in solution is a dynamic ensemble of the R and R2 structures and that the deoxy HbA in solution is different from all known high resolution crystals structures available so far and this has led to the suggestion that multiple quaternary structures of hemoglobin, including the end-state-ligated conformations and intermediate structures, may coexist under physiological conditions, each of them contributing to the average quaternary structure determined as the solution structure. The structures obtained from crystallography studies provide only static information of the protein before and after ligand binding, but not on the dynamics of the transition. Therefore, the existence of different crystal structures of the fully ligated HbA has raised questions on whether transition from the unligated structure to the fully ligated structure proceeds via intermediate forms or that the fully ligated hemoglobin is actually an ensemble of energetically accessible structures that are in dynamic equilibrium under physiological conditions as suggested from the NMR studies. It has also been reported that the functional properties of HbA in crystals are different from those in solution, for example, the absence of a Bohr Effect, cooperativity, and the effect of allosteric effector in the Hb crystals (14-16).

Computer simulation techniques, and especially molecular dynamics (MD) simulations, which can provide a fully atomistic description of the dynamical features of a given system, are well suited to complement the experimental studies on the conformational transitions in hemoglobin. However, because of hemoglobin's large size, early computer simulation studies on HbA have either been restricted to only a part of the protein (17-20), or relatively short time simulations for the entire protein (21-24). More recently, longer simulations of hemoglobin on nanosecond time scale became possible (25-29). In particular, Hub *et al* (28) performed long simulations on the quaternary transitions for hemoglobin x-ray structures. However, their study is limited to the deoxy- form, and needs to be complemented with studies of the oxy- form. Here we report a long-time scale full atomistic MD simulation study on three different x-ray structures and a NMR structure of HbA in both deoxy- and oxy- forms, to probe the global dynamics and the conformational changes that occur in the molecule. Our work shows that the R structure is more stable than the T structure irrespectively of the presence of the oxygen, and clarifies the role of the subunit interactions in the induced quaternary structure transition.

## METHODS

### Initial set up

The starting structures for all the simulations and comparison were taken from the Protein Data Bank (PDB). The x-ray structures with the PDB code 2HHB (30), 1HHO (31), 1BBB (7), 1MKO (9) and 1YZI (9) were used as the T, R, R2, RR2 and R3 conformations respectively. A simulation was also carried out starting from a NMR structure of deoxyhemoglobin, PDB code 2H35 (model 1) (32). Simulations were performed for four structures in both the unligated deoxy- and the fully-ligated oxy-forms. Therefore a total of 8 structures were simulated; T-state deoxygenated and oxygenated HbA (T-$noO_2$ and T-$O_2$), R-state deoxygenated and oxygenated HbA (R-$noO_2$ and R-$O_2$), R2-state deoxygenated and oxygenated HbA (R2-$noO_2$ and R2-$O_2$), and the NMR-deoxygenated and oxygenated HbA (N-$noO_2$ and N-$O_2$). The R2-$O_2$ structure was obtained by replacing the CO in the PDB structure with $O_2$ while the T-$O_2$ and N-$O_2$ structures were obtained by adding $O_2$ to their PDB structures at an appropriate bonding distance from the iron atoms of the hemes. All of the histidines were protonated at the ε-position except the proximal histidines, which were protonated at the δ-position since the ε-position is used to connect to the heme. The deoxy- and oxy- forms of the starting structures were explicitly solvated in cubic box of water of volume 23301 Å$^3$, 25738 Å$^3$, 27905 Å$^3$ and 24812Å$^3$ for the T, R, R2 and NMR structures respectively. The solvation box has a density of 1.0 gcm$^{-3}$ and a net charge of -6. Six sodium counter ions were added to neutralize the system.

### Energy Minimization and MD simulations.

All the simulations were carried out using the GROMACS simulation software (33-34). The Amber force field 99-SB (35) and the TIP3P water model (36) were used. The parameters for oxygenated heme group obtained according to the standard Amber force field protocol by Estrin's group (36) was made available for this work. The LINCS (37) algorithms was used to constrain all the bonds, the electrostatic interactions were computed at every step using the particle-mesh Ewald summation algorithm (38-39). Short-range repulsive and attractive interactions were described by a Lennard-Jones potential within a cut off of 10.0 Å. The steepest descent followed by conjugate gradient method was used for the energy minimization. The bulk water was first equilibrated for 20 ps keeping the protein atoms positions fixed to allow for the relaxation of the water molecules. The entire system was minimized again and then subjected to 500 ps equilibration by gradually heating up system from 0-300 K at 20 K intervals. The MD simulations were carried out at constant temperature and pressure (NPT) of 300 K and 1atm for 100 ns with time-step of 2 fs. Simulations were performed at ambient conditions by coupling the system with the Nose-Hover (41) thermostat and the isotropic Parrinello-Rahman (42) barostat.

## RESULTS AND DISCUSSION

### Global dynamics of Hb molecule.

HbA structural changes in both the deoxy- and oxy- forms for the T, R, R2 and NMR structures during the MD simulation were followed by monitoring the root-mean square deviation (RMSD) with respect to the starting structures of the simulation (Fig.1).

The removal of the ligand from the R and R2 structures is expected to initiate the R→T transition. However, no quaternary structural changes were observed during the 100 ns time window spanned by our simulations. In this time frame, the protein atoms in the relaxed R-conformations tend to maintain their overall stability irrespective of the presence (R-$O_2$ and R2-$O_2$) or absence of the ligand (R-no$O_2$ and R2-no$O_2$), as it is evident from the protein backbone RMSD. A similar result is obtained starting from the NMR structure (N-no$O_2$ and N-$O_2$). The major difference between the R and R2 structure and the N structure is that the latter shows a larger RMSD from the initial structure with more pronounced amplitude of the atomic fluctuations. The NMR structure of deoxy HbA in solution seems to be markedly different from the x-ray T and R structures and not along the T→R transition path. On the other hand, the simulation starting from the T-conformation, T-no$O_2$ and T-$O_2$ exhibited a quaternary structural transition as evident from the RMSD plots (Fig. 1) and RMSD matrix for the backbone atoms (Fig. 2). In particular, the RMSD matrix (Fig.2) shows the transition between two distinct conformational states, *i.e.* the starting T-conformation and the final R-conformation. No such transitions are present in the other cases.

The calculation of the RMSD from the trajectory for the T structures with respect to the various ligated x-ray structures (R, R2, RR2 and R3) indicates that the observed structural change corresponds to the T→R transition (Fig.3). From the plots reported in Fig. 3, it is evident that this transition leads to a structure closer to R than R2, RR2, and R3. Indeed, the average RMSD values calculated on the backbone atoms of the T-$O_2$ structure from our MD simulation are 1.62 Å , 2.50 Å, 1.97 Å and 2.38 Å as against the experimental values of 2.33 Å , 3.48 Å, 3.05 Å and 3.80 Å for R, R2, RR2 and R3 x-ray structures respectively. Similar conclusion is drawn from the super-imposition of x-ray and MD T-structures on the x-ray R-structure, which shows that the quaternary structure of both T-$O_2$ and T-no$O_2$ are R-like after the MD simulation (Fig.4). The present results are consistent with earlier MD simulation studies. For example Hub *et al* (28) reported similar results for the deoxy-forms of T, R and R2 conformations. Also, Laberge and Yonetai (26) performed short MD study that, in spite of not being long enough to sample quaternary transition, showed that the T-conformation is dynamically more flexible than the R conformation and that its fluctuations are perturbed by the presence of DPG.

Remarkably, no structural change is observed in the RMSD if we use as reference the experimental NMR deoxy conformation (Fig. 5b). Indeed, the RMSD calculated for N-$O_2$ and N-no$O_2$ simulations with respect to both the T and R structures with respect to the NMR structure (Fig.6) does not show any appreciable difference. This finding suggests that the NMR structures might not be along the T→R transition path.

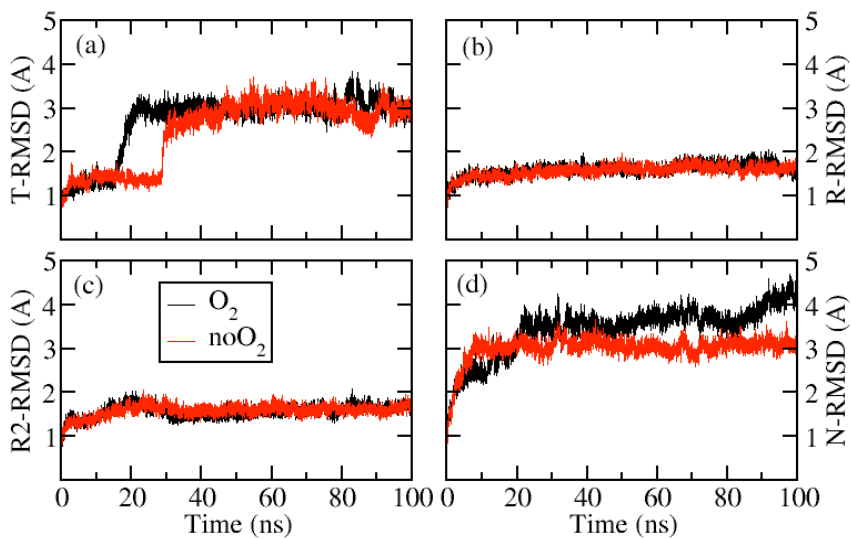

***Fig.1**(a-d): RMSD plots showing the dynamic motion of the backbone atoms during the simulation time when compared to the starting structure of the simulation. (a) T-noO$_2$ and T-O$_2$ (b) R-noO$_2$ and R-O$_2$, (c) R2-noO$_2$ and R2-O$_2$, (d) N-noO$_2$ and N-O$_2$. Only the T conformation shows quaternary structural transition*

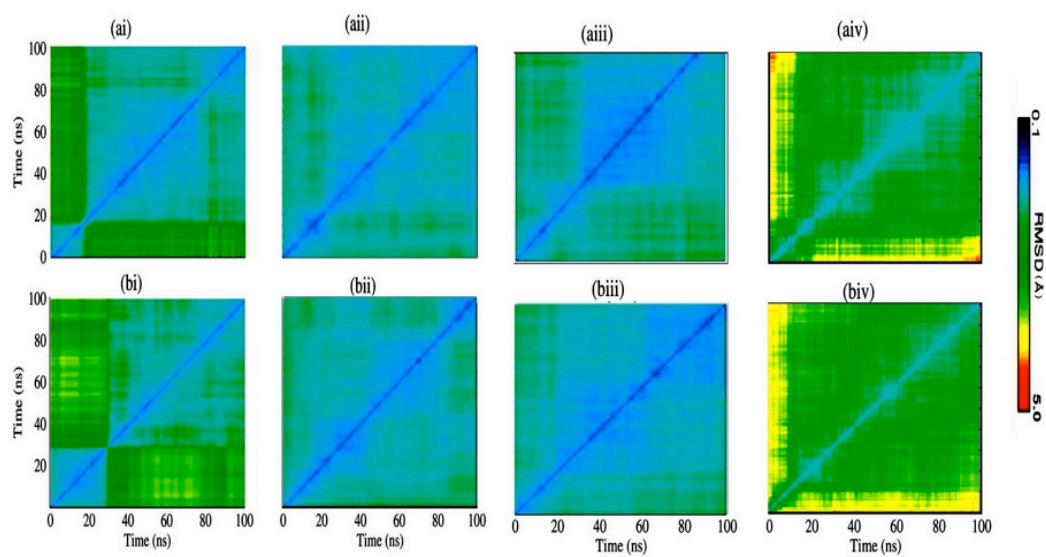

***Fig.2**:ai-iv, RMSD matrix plots for the oxy- structures; T-O$_2$, R-O$_2$, R2-O$_2$, and N–O$_2$ respectively, whereas bi-iv, is the corresponding matrices for the deoxy- structures; T-noO$_2$, R-noO$_2$, R2-noO$_2$ and N-noO$_2$. In both (a) and (b) only the T-conformation shows two distinct structural states representing the T→R transition.*

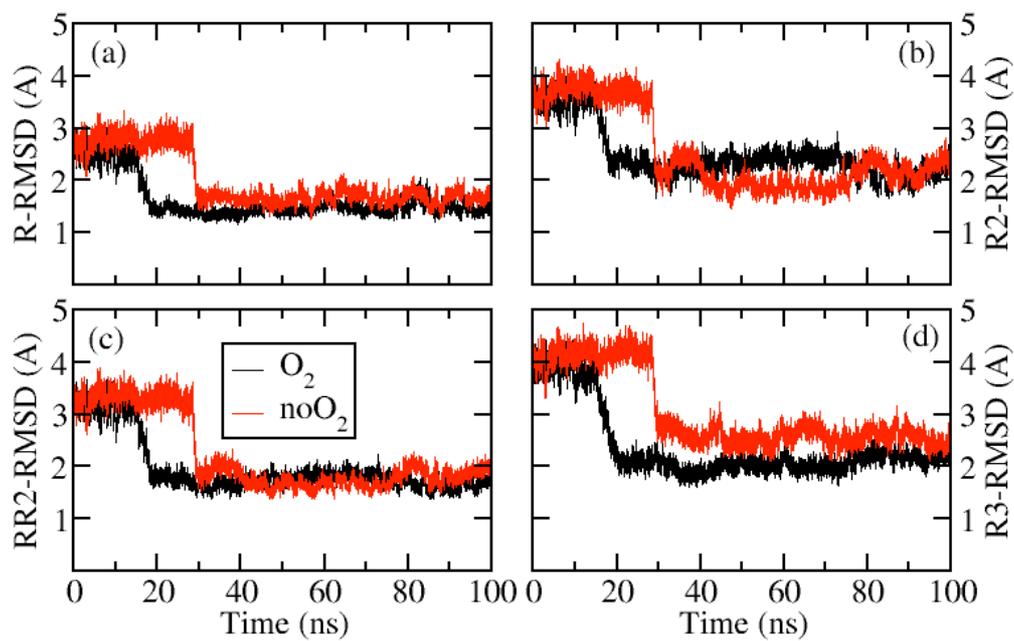

***Fig.3a-d:*** *RMSD plots of the MD trajectories of the oxy- and deoxy- T conformation when compared to the different experimental ligated (R (panel a), R2 (panel b), RR2 (panel 3) and R3 (panel 4)) X-ray structures*

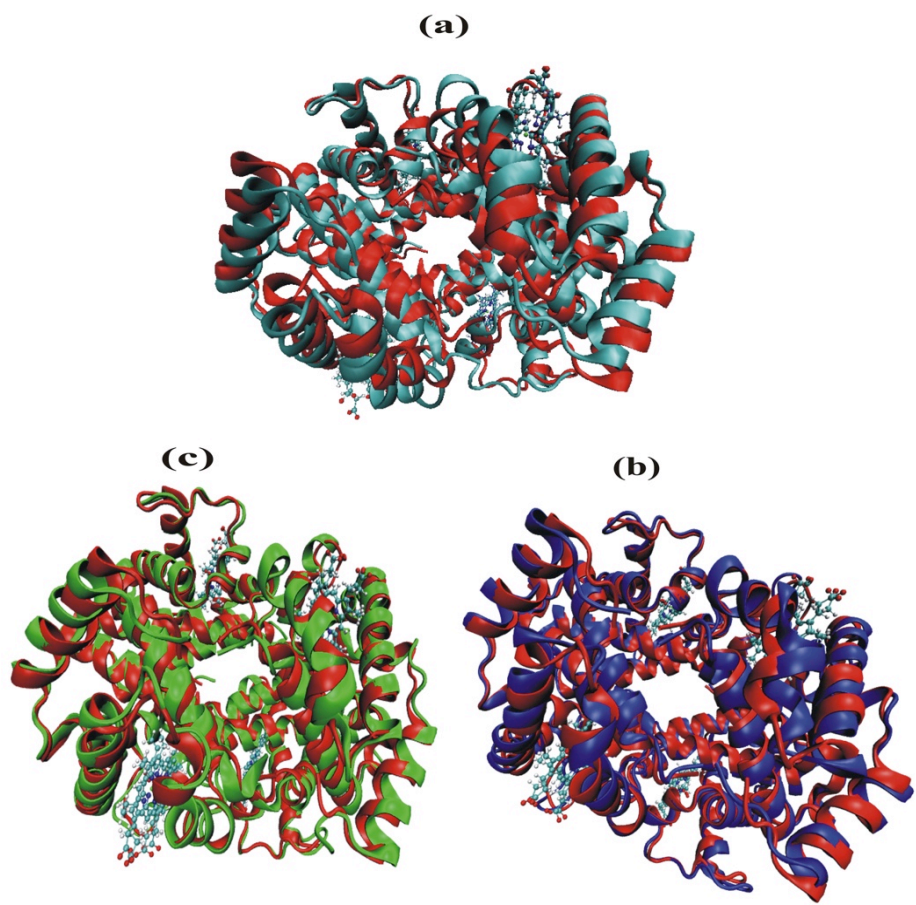

***Fig. 4a-c****: (a) Superimposition of the T (cyan) and R (red) structures as obtained from Xray. (b) Superimposition of the final structure of simulations T-noO2 (panel b, blue) and T-O2 (panel c, green) with the R (red) structure as obtained from X-ray .*

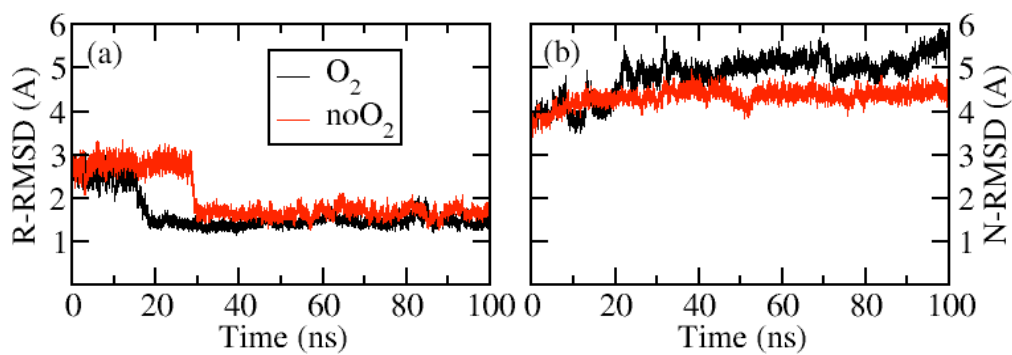

**Fig. 5a-b:** *RMSD plots of the MD trajectories of the oxy- and deoxy- T conformation when compared to (a) the experimental R X-ray structures and (b) the NMR solution structure. When compared with the NMR structure, the trajectory does not show any structural transition.*

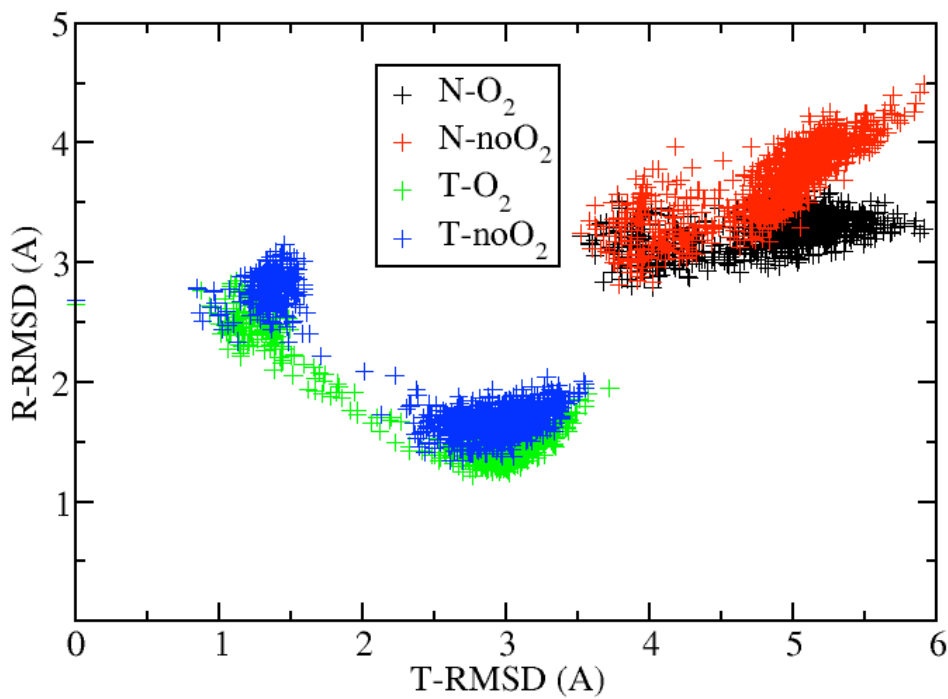

**Fig.6:** Root mean square displacement with the T experimental structure (horizontal axis) and with the R experimental structure (vertical axis) of the N-$O_2$, N-no$O_2$, T-$O_2$ and T-no$O_2$ simulations. It can be clearly seen that the NMR structure does not lie along the T-R path.

**The Role of the subunits in T→R Quaternary transition.**

Changes in the tertiary structure were analyzed in terms of the RMSD calculated for the α- and β- subunits with respect to the corresponding x-ray structures of the two subunits. The overall structural change of the individual α- and β- subunits during the simulation was not so significant except for the residues at the amino- and carboxy-termini and those at inter-dimer α-β interfaces ($α_1β_2$, and $α_2β_1$, see Fig. 7). We observed dynamical fluctuations of both the α- and β- subunits being the amplitude of the fluctuations of the β-subunits larger. Consequently the β- subunits have a more pronounced displacement from the starting structure (Fig.8a). There are also notable structural changes at the hemes in both α- and β- subunits during the simulations. In the X-Ray T-structure, the porphyrin ring of the hemes is domed and the iron atom is displaced from the plane of the porphyrin nitrogen atoms towards the proximal histidines. However, during the MD of the T-$O_2$ and T-no$O_2$, the porphyrins flatten and the iron atom move toward the plane of the porphyrins resulting in ≈ 0.5 Å increment in the Fe-$N_\varepsilon$-HisF8 bond (Fig 9). Overall, the tertiary conformational transition is not as evident as the quaternary transition is. The motion of the terminal residues (Arg 141 of the α-subunits and His 146 of the β-subunits) is essentially more pronounced, underscoring the role of these residues in the quaternary transition (Fig. 10). This motion of the terminal residues resulted in the delocalization of the adjoining residues (Tyr 140α and Tyr 145β) and the bending of the hydrogen bonds formed by Tyr 140α – Val 93α and Tyr 145β – Val 98β . Earlier studies on the T- and R- x-ray structures (2,3,5-6) have indicated that the tertiary structural change are only significant in localized regions which include the hemes, the F helices, FG corners of both α-,β-subunits and the E helix of β subunits. Similar observation was also reported from the NMR study of deoxyhemoglobin in solution which shows that there are only subtle differences at the tertiary structure levels between the deoxy- and carbonmonoxy- structures of HbA (12), and consequently; the tertiary structures of T and R-states of HbA are only slightly different. In summary, the tertiary structural change of each subunit is not large and the internal motion of the four subunits cannot be properly used to detect the quaternary transition unless the motions of all the subunits are coupled together. This could indicate that when all the subunits are coupled together this small change in the tertiary structure is reflected in a large change of the quaternary structure.

**The Role of the subunit contacts in T→R Quaternary transition.**

The dynamics of the residues at the inter-subunits contacts plays important roles in the structural changes occurring in hemoglobin during the quartenary transition. The inter-subunit regions where there are possible contacts between the four subunits of HbA are the $\alpha_1\beta_1$, $\alpha_2\beta_2$, $\alpha_1\beta_2$, and $\alpha_2\beta_1$ interfaces (Fig. 7). The movement of the residues at the $\alpha_1\beta_2$ and $\alpha_2\beta_1$ interfaces has been employed to explain the mechanisms of quaternary transition (the so-called quaternary-switch) (3). In addition, the interactions at the inter-subunits interfaces have been reported to induce tertiary structural changes and changes in the heme pocket of HbA(43). Here we analyze the dynamics of the inter-subunit contacts. In the ensuing discussion, we have used a cut-off of 10 Å to define the interfacial residues in the x-ray T-structure.

Our simulations indicate that, in both the deoxy and the ligated structures, the $\alpha_1\beta_1$ and $\alpha_2\beta_2$ interfaces are relatively rigid and, consequently, are not involved in the observed transition as evident from Fig. 8b. This is consistent with structural similarity between the x-ray structures of deoxy and ligated HbA at these interfaces (2, 3). Since these regions are rigid, they can be taken as reference frame to describe the conformation change between T and R structures.

The quaternary-switch is clearly evident in our MD simulations of the T-structures and it corresponds to the observed T→R transition Fig.8b. This interface is made up of all of the αC, αFG, βC, βFG, residues, the first three residues of αG and βG and the last two residues the αHC and βHC corners. The quaternary transition involves the relative rotation $\alpha_1C\beta_2FG$ residues at $\alpha_1\beta_2$ interface and the $\alpha_2C\beta_1FG$ residues at the $\alpha_2\beta_1$ interface. This rotation results notably, in the global movement of the Hb molecule, the relative change in the position of His 97β from its initial position between Pro 44α and Thr 41α by a complete turn to another position between Thr41α and Thr38α (Fig. 11), the breaking of salt bridges formed by the terminal Arg141α and His146β residues and the H-bonds at the $\alpha_1\beta_2$ ($\alpha_2\beta_1$) interface between the Asp99β and Tyr42α and Asn97α residues in the T structure among other changes. These observations are in agreement with earlier reports (2-3). The T→R conformational transition observed from our simulations starting from the T-structure in both the deoxy-and oxy forms may be connected with the fact that, under physiological conditions, the equilibrium between T- and R- forms of Hb is regulated by allosteric interactions and by the effects of heterotropic effectors like $H^+$, $Cl^-$, $CO_2$ and organic phosphate groups like 2,3-bisphosphoglycerate (2,3-BPG) and inositol hexaphosphate (IHP). The deoxy-T conformation is stabilized both by low pH and the binding of these heterotropic ligands at different positions of the protein and the release of these effectors on oxygenation of Hb is necessary for the T↔ R equilibrium to shift in favor of the R conformation and vice versa. The unexpected transition of the deoxy T conformation to R observed from our MD simulation could be attributed to the fact that the dynamical interactions among the four subunits in Hb without the effect of the heterotropic ligands favors the 'relaxed' R-conformations and such responsible to observed T→R transitions observed in both oxy- and deoxy-T structure simulations.

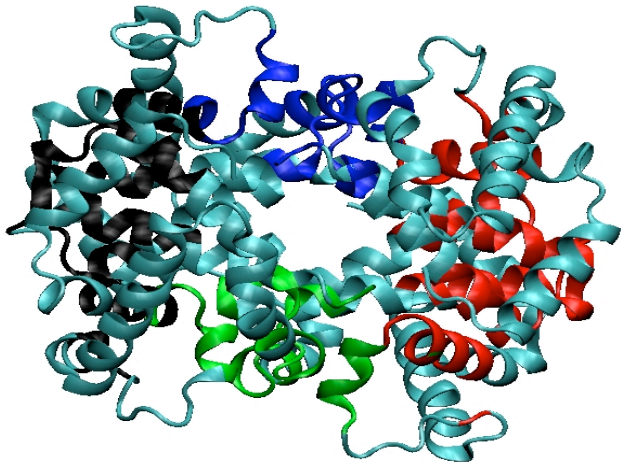

***Fig.7***: *The structure of HbA showing the inter subunit interfaces; the black color shows the $\alpha_1\beta_2$ interface, the red color is for the $\alpha_2\beta_1$ interface, the green color for the $\alpha_1\beta_1$ interface and the blue color for the $\alpha_2\beta_2$ interface.*

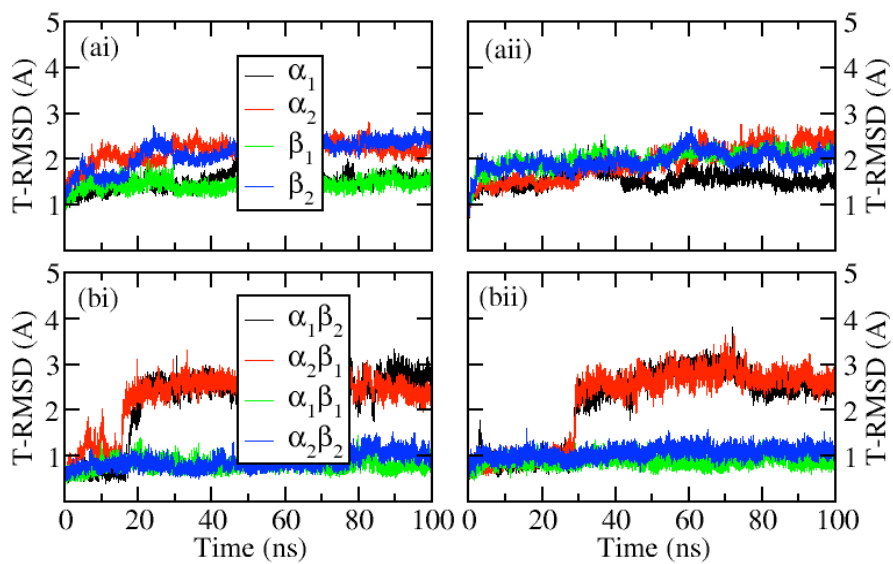

***Fig.8****: RMSD plots with respect to the starting structure for the four subunits in T-$O_2$ and T-no$O_2$ structures during the simulation (ai and aii), and the RMSD plot with respect to the starting structure for the interfaces T-$O_2$ and T-no$O_2$ structures (bi-bii).*

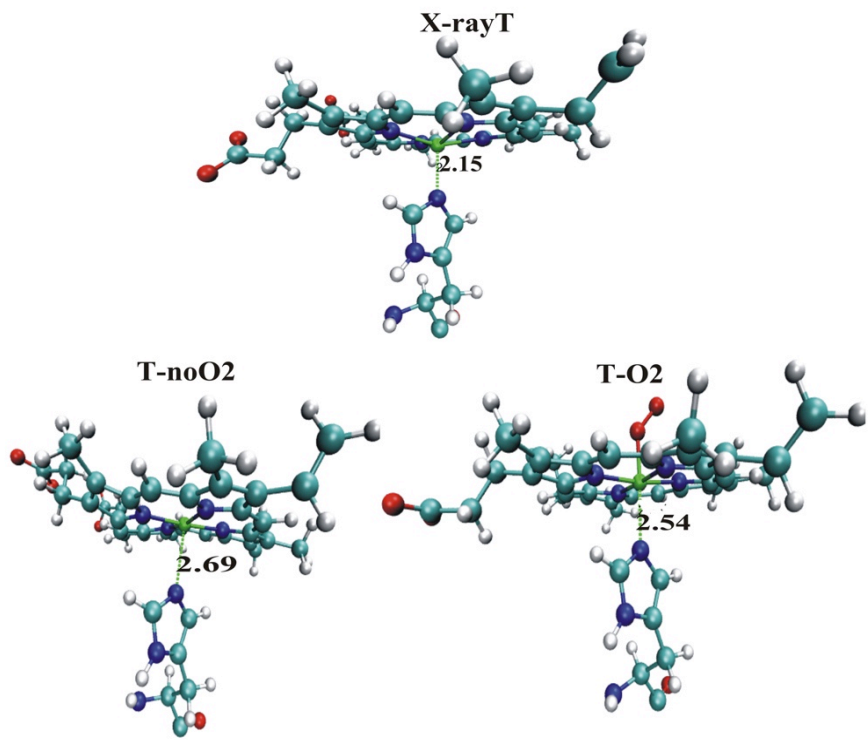

**Fig. 9**: Structure of the heme with the proximal histidine in the x-rayT and the MD T-$O_2$ and T-no$O_2$ structures.

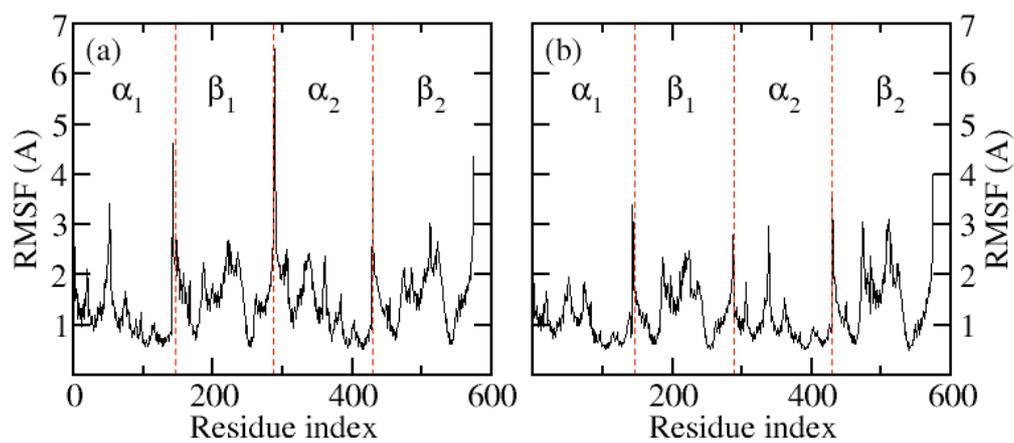

**Fig. 10**: *The RMS fluctuation plot of the $C_\alpha$-atoms of (a). T-noO$_2$ and (b). T-O$_2$ structures during the MD simulation, the terminal residues of all the four subunits have the largest RMS fluctuation.*

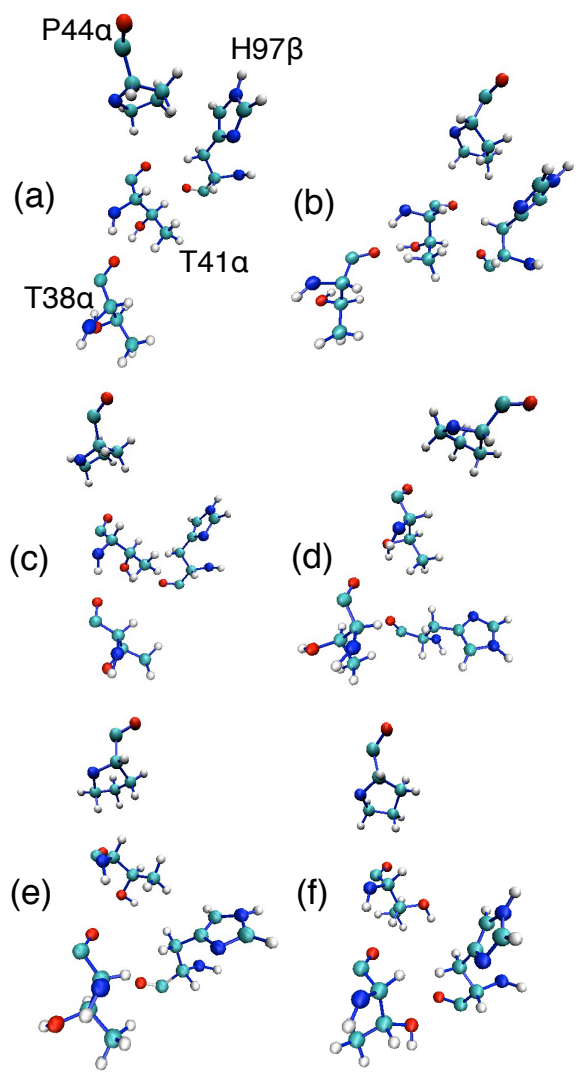

Fig.11 : Snapshots of residues His 97β, Thr 41βα, Thr 38α and Pro 44α taken along the simulation of T-O2 and from experimental structures. (a) X-ray (T structure), (b-e) MD simulation of T-O2 at 10 ns (b), 20ns (c), 30 ns (d) and 100 ns (e). (f) X-ray R-structure. As can be seen, His 97β moves from between Pro 44α and Thr 41α to between Thr 41α Thr 38α during the MD simulation of T-O2, consistent with the experimental R structure.

**CONCLUSION**

The quaternary conformation of HbA switches between the 'tensed' or T-conformation and the 'relaxed' or R-conformation(s). We have demonstrated through our MD studies that quaternary structural change of hemoglobin in the T conformation is accessible within ns timescale and that the T→R transition and it is largely dependent on the dynamical interactions between the four subunits, especially the residues at the inter-dimer interface. The R conformations are relatively more rigid when compared to T conformation and, as such, the R→T quaternary transition was not observed on the sub micro-second timescale span by our simulations. Further investigations are deemed necessary to assess the thermodynamic stability of one conformational state with respect to another. Our study also confirms the that the NMR solution structure of HbA is different from all the x-ray structures conformations, thus, further studies are necessary that describe the complementary relations between the structures of HbA in solution determined by NMR studies and the ones from x-ray crystallographic studies.

**AKNOWLEDGMENT**

The authors wish to acknowledge the financial support of ICTP/IAEA Sandwich Training Education Programme (STEP),  and Prof. D. A. Estrin for making available to us the oxyheme force field parameters from his group.